%% file: main.tex
\documentclass[a4paper, amsfonts, amssymb, amsmath, reprint, showkeys, nofootinbib, twoside]{revtex4-2}
\usepackage[english]{babel}
\usepackage[utf8]{inputenc}
\usepackage[colorinlistoftodos, color=green!40, prependcaption]{todonotes}
\input{preamble}
\usepackage[pdftex, pdftitle={Article}, pdfauthor={Author}]{hyperref} 
\usepackage{empheq,etoolbox}
\patchcmd{\subequations}
  {\theparentequation\alph{equation}}
  {\theparentequation.\alph{equation}}
  {}{}
\begin{document}
\title{Harmonic fractal transformation, 4R-regeneration and noise shaping for ultra wide-band reception in FitzHugh-Nagumo neuronal model}

\author{Mariia Sorokina}
    \email[Correspondence email address: ]{k4939b@gmail.com}
    \affiliation{90 Navigation street, Birmingham, UK, B5 4AA}

\date{\today} 

\begin{abstract}
Human hearing range significantly surpasses the typical neuronal spiking frequency. Yet, neurons with their modest frequency range not only efficiently  receive and process multiple orders higher frequency signals, but also demonstrate remarkable stability and adaptability to  frequency variations in brain functional connectivity. Ability to process signals beyond the limitations of the receiver temporal or frequency (bandwidth) resolution is highly desirable yet requires complex design architectures. 
Using the FitzHugh-Nagumo model we reveal   the harmonic fractal transformation of frequency and bandwidth, which enables the Nyquist rate integer (for low frequencies) and sub-integer (for high frequencies) multiplication.
  We also demonstrate  for the first time that noise shaping can be achieved in a simple RLC-circuit  without a requirement of a delay line. The discovered effect presents a novel regeneration type - 4R: re-amplifying, re-shaping, re-timing, and re-modulating and due to the fractal nature of transformation offers a remarkable regenerative efficiency.   The effect is a generalization of phase locking to non-periodic encoded signals. The discovered physical mechanism explains how using neuronal functionality one can receive and process signals over an ultra-wide band (below or higher the spiking neuronal range by multiple orders) and below the noise floor.  
\end{abstract}


\maketitle

\section*{Introduction}
Any receiver is limited by its resolution. While,  the receiver must sustain a minimum sampling rate - twice the signal bandwidth - for zero-aliasing in signal reconstruction, i.e. the Nyquist rate \cite{Shannon, Nyquist}. There is a growing interest in capturing ultra-fast dynamics or detecting signals below the noise floor, which require complex design architectures  \cite{TS}. Also, modern communication systems aim to widen the frequency range (e.g. 5G and 6G), which sets the bandwidth requirements for neuromorphic processing  \cite{THz,MC}. 

While it is clear that a biological neuron as a physical system has resolution limitations in signal reception and processing, yet the problem of establishing these limits in the neuronal and neural models remains unaddressed. Furthermore, the mechanism of processing wide reception bands characteristic to animal hearing curves \cite{ATH} remains unexplained.  

Since the seminal work on the frequency multiplication effect in the Van der Pol oscillator \cite{VdP} has led to the discovery of the phase locking effect in neuronal systems, the subsequent research on this topic was focused on the analysis of periodic properties instead of spectral. While noise shaping is commonly used in data processing for  improving signal quality  \cite{NS},  recently, noise shaping has been observed in neural networks \cite{ekchorn}, thus calling for more detailed revision of spectral effects in neural systems.

Here we unveil the harmonic multiplication effect based on the FitzHugh-Nagumo (FHN) model  \cite{FHN1,FHN2}.  The resulted harmonic fractal transformation (HFT) multipliers  a signal frequency (for periodic input) or bandwidth (for encoded input)  adjusting it to the FHN spiking frequency, thus, enabling reception of multiple orders larger or smaller input frequencies enabling the Nyquist rate integer and sub-integer multiplication augmented by the noise shaping. Moreover, we extend the studies of phase locking to communication signals, which are non-periodic and randomly encoded. 
Overall, here we demonstrate and explain how neuron-like spiking enables processing signals beyond the receiver resolution limits with an additional advantage of noise suppression. Moreover, we demonstrate the noise shaping without any dithering or delay lines.  The results reveal a new type of signal regeneration.
The fractal nature of transformation (a self-similar and infinitely repeated pattern) enables a frequency quantization, while  preserving the input characteristics at the infinitely diminishing self-similar scale. Moreover, similarly to the advantages of analogue computing in neuromorphic processing, we illustrate  the advantages of semi-analogue quantization on the example  of the FHN circuit for neuromorphic communications.

\section*{Results}

\begin{figure*}[!ht]
\centering
\includegraphics[width=\linewidth]{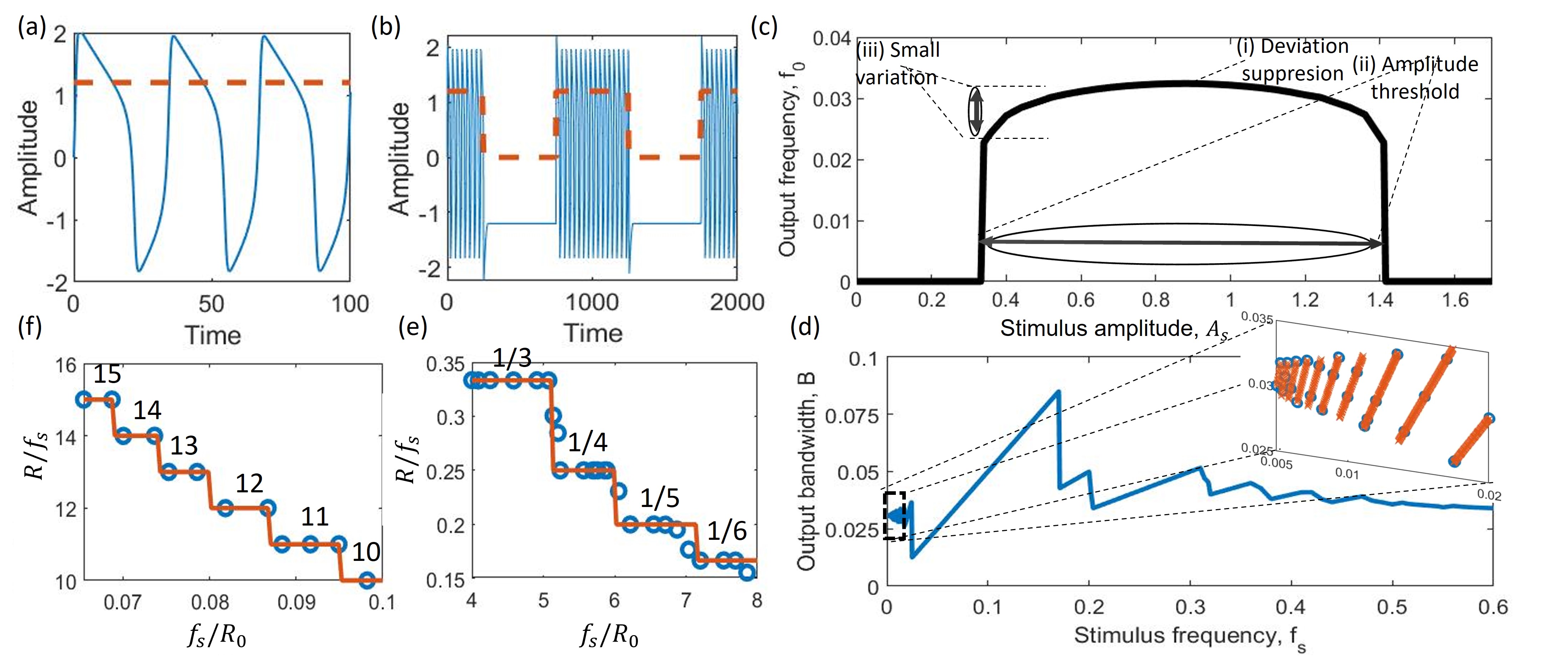}
\caption{\textbf{Bandwidth and Nyquist rate.} The FHN output (blue solid lines) for a stimulus (red dashed) with within-the-threshold amplitude $A_s=1.2$: \textbf{a}  constant and \textbf{b} modulated with symbol frequency $f_s=0.001$.  \textbf{c} The output frequency $f_0$ for a constant stimulus as a function of its amplitude $A_s$. The conditions for 4R-regeneration: (i) deviation suppression; (ii) amplitude threshold; and (iii) small frequency variation within the threshold region.  \textbf{d} The output bandwidth $B(f_s)$ as a function of stimulus symbol frequency with fixed $A_s=1.2$. The output bandwidth is preserved to the eigen-value $B\rightarrow f_0(A_s)$ (see panel c) with the precision increasing for larger deviations (when $f_s\ll f_0$ and $f_s\gg f_0$). This is due to the fractal nature of the transfer function $B(f_s)$, see enlargement of $f_s\ll f_0$ in the inset (with the analytical result $f_s[R_0/f_s]$ ($R_0=2f_0$) plotted in red).  The output Nyquist rate $R=2B$ is an \textbf{integer sub-multiple} or an \textbf{integer multiple} of input symbol frequency $f_s$: plotted for $f_s>f_0$ and $f_s<f_0$ in panels \textbf{e}  and \textbf{f}  in blue alongside the analytics in red.  }
\label{F1}
\end{figure*}

The Fitzhugh-Nagumo (FHN) model \cite{FHN1,FHN2}: 
\begin{subequations}
  \begin{empheq}[left=\empheqlbrace]{align}
\dot{v} &=v-\frac{v^3}{3}-w+I \\
\tau \dot{w} &=v-bw+a
  \end{empheq}
  \label{FNH_o}
\end{subequations}
was proposed as one of the simplest equations to model neuronal spiking (see Fig. \ref{F1}a) with an amplitude threshold, outside which the excitation block occurs. Here $I$ represent an input stimulus and the parameters are further fixed as $a=0.8, b=0.7, \ \tau=10$, unless otherwise specified. For modulated stimulus bursting  \cite{review,LA}    (see Fig. \ref{F1}b) or high frequency blocking \cite{HF}, similarly to the biological neurons, may occur. 

We note that the FHN node acts as a modulator, which combines a frequency-shift keying and an amplitude-shift keying in a nonlinear way:   a \textit{constant} stimulus with amplitude (see Fig. \ref{F1}a) excites an output modulated with frequency $f_{0}$ as a step-like function of the stimulus amplitude  (see Fig. \ref{F1}c), see spectrum analysis in Supplemental Note I.A. 
 Moreover, one can observe that the transfer function (TF), see Fig. \ref{F1}c),   $f_0(A_s)$ exhibit the following properties: (i) regeneration condition (i.e. deviation suppression) $f_0'(A_s)\leq 1$ (see derivations in Supplemental Note I.B.); (ii) amplitude threshold region, within which the condition (i) is fulfilled.  This is the first example of regenerative functionality in a spectral domain. Moreover, the condition (ii) enables a novel regeneration type - regeneration inacts only within the power threshold (enabling in-built energy efficiency).

We show that for a modulated input a harmonic locking occurs: the spectral components (spikes) at the harmonic closest to $f_0$ are amplified (see Supplemental Note II.A.). As a result, the amplified spike represents the largest frequency component and, consequently, bandwidth. Thus, the output bandwidth is preserved via locking the symbol frequency ($f_s$) harmonic term that is closest to the eigen-frequency $f_0$, as a result the bandwidth is preserved to the value of the eigen-frequency $f_0$. See in Fig. 1d) the output bandwidth  for On-Off Keying (OOK) encoded stimulus with varied symbol frequency $f_s$ and fixed amplitude $A_s=1.2$. 
As locking mechanism is related to the harmonic terms, the spectral spike may occur at frequency components $nf_s$ (for periodic signals, Supplemental Note II.A.) and in-between  $(n\pm 1/2)f_s$ (for encoded signals, see Supplemental Note II.B.). The latter case  is due to the due to the resonance beating of two harmonics enables by the continuous nature of the encoded input stimulus   spectrum. 
The inset in Fig. 1d) highlights the case of slow-varied encoded stimuli $f_s\ll f_0$:   the output bandwidth follows the nearest integer function (rounding half up) $[2f_0/f_s]$; while, for uncoded periodic stimuli the output frequency will be given by $[f_0/f_s]$. These conditions (regenerative transfer function for varied stimulus amplitude and harmonic locking for varied stimulus frequency) enable the bandwidth of the generated randomly encoded signals to be \textit{preserved to the FHN eigen-frequency for various stimulus frequencies or modulation waveforms} (see Fig. 1d) for OOK, while various sinc and rrc pulses are shown in Supplemental Note II.B.). This is the first observation of harmonic locking in neuronal models.

In other words, the FHN performs the \textit{integer factorization of the Nyquist rate}. See the corresponding Nyquist rate $R=2B$ ratio to the input symbol frequency $f_s$ shown in Fig. 1e) in blue alongside the analytical formula  $f_s[R_0/f_s]$ in red (with the octave $R_0=2f_0$ further referred as the \textit{eigen-rate}). While, for large frequency $f_s\gg f_0$ the output can be approximated as $f_s/[f_s/(R_0(1+1/[f_s/\hat{R}])]$ with $\hat{R}=R_0(\langle A_s\rangle)$ (see Supplemental Note II.B.) as shown in red solid lines. 
 This is the first demonstration of the Nyquist rate integer factorization in neuronal models and in  a single RLC circuit without a delay line (the FHN circuit). This is a new type of regeneration - 4R: re-amplifying, re-shaping, re-timing, and re-modulating.

\begin{figure*}[!ht]
\centering
\includegraphics[width=\linewidth]{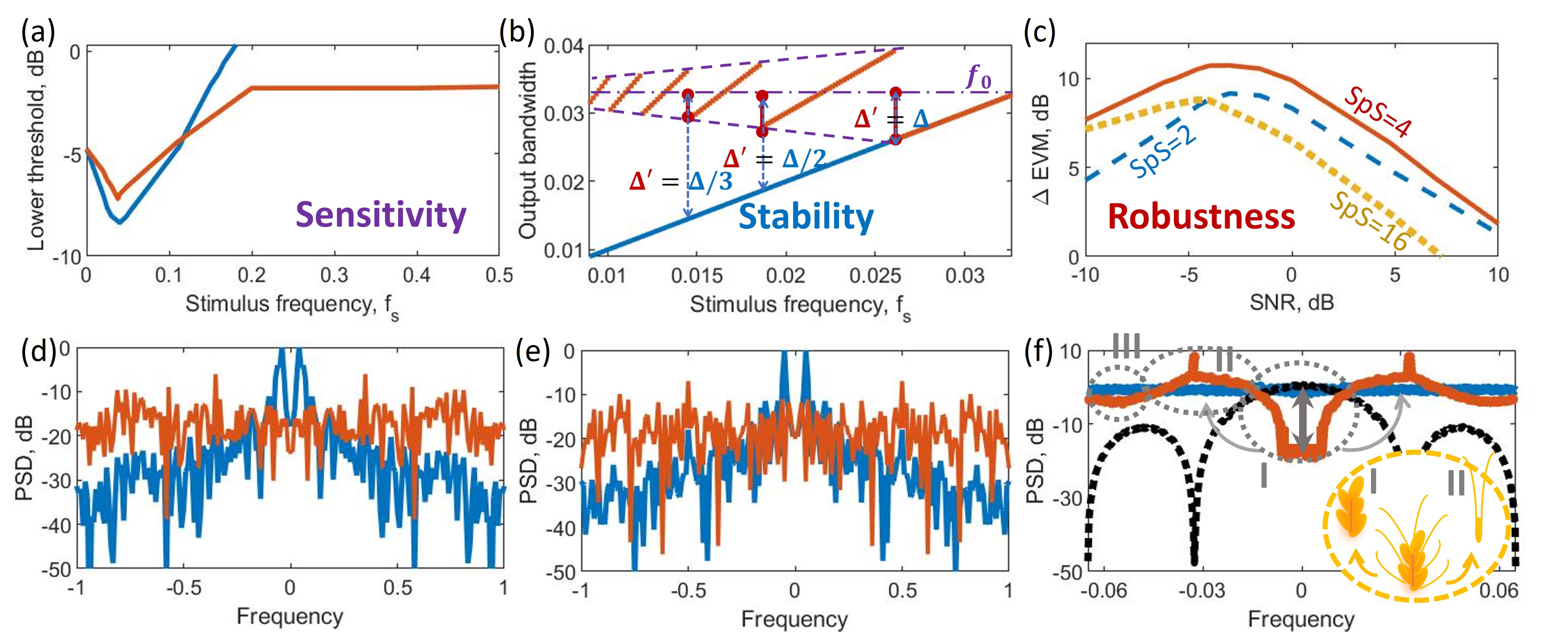}
\caption{\textbf{Impact: } \textbf{a)} \textbf{frequency sensitivity} -  the lower amplitude threshold for  pure-tone (blue)   and shifted (red)  sine-wave stimulus as a function of frequency $f_s$ demonstrates high  frequency sensitivity (analogous to the absolute threshold of hearing effect). 
 \textbf{b)} \textbf{bandwidth stability} -
 the output bandwidth (red) is harmonic fractal transformed (linear transformation is shown in blue) to the eigen-value $f_0$ resulting in  \textit{higher precision for bigger deviation}: compare the input deviation in blue $\Delta$ and the output deviation $\Delta'$ in red (the upper bound on the output deviation is shown by violet dashed lines);
\textbf{c)} \textbf{robustness to noise} - the FHN induced noise shaping improves the EVM  (compared to the corresponding oversampled linear case).
Assuming a receiver with fixed $\mathrm{RBW}=0.01$ the input signals (shown in red) with $f_s=0.001$ \textbf{d)}  and $f_s=0.5$ \textbf{e)} below the noise floor can be detected after the FHN transformation  (blue).  
\textbf{f)} For rectangular modulated encoded input (black, dashed) contaminated with the uniformly distributed noise (blue), the FHN transformation reshapes the noise (red) moving it from the 3dB bandwidth area (denoted as I) to  higher frequencies (II) in repeated patterns (III).  The inset shows the grain packing analogy:  aligned and rotated grains (I) and the awns (II).}
\label{F2}
\end{figure*}

The frequency transformation $\mathrm{FHN}(f_s)\rightarrow f_0$ results in the exceptional \textit{sensitivity} in detection as the FHN adjusts signal frequency to the desired value ($f_0$) in a broad frequency and amplitude range. 
Note, that for stimulus modulated at high frequencies $f_s>f_0$ the waveform becomes important, note in Fig. 1e) the narrow frequency intervals where the violations of phase locking occurs: the intervals increase as the stimulus frequency grows.  In our example we consider a pure tone sine wave $A_s \sin(2\pi tf_s)$ (blue in Fig. 2a))  and a shifted positively defined sine wave $A_s/2[1+sin(2\pi tf_s)]$ (red).  Apart for the same frequency $f_s$, both waveforms share the same maximum stimulus amplitude $A_s$, minimizing which we can estimate the lower amplitude threshold. The difference is that the shifted wave is more localized near the balancing value $A^*=a/b$, thus contributing more energy to spiking. 

In Fig. 2a) we show that the lower amplitude threshold - the lowest amplitude, which generates spiking, has a high passing range in frequency. This means that signals with very low or very high frequencies still excite spiking. There are two  conclusions from Figure 2a):
\\
(1)   \textbf{ stimulus frequency range is multiple orders wider}  compared to the spiking range;\\
(2) the amplitude threshold is little varied over the frequency range.

Compare the results plotted in Fig. 2a) with the typical absolute threshold of hearing (ATH) curve for a human\cite{ATH}:
\\
(1) the sound perception varies from 0.1 to 10 kHz, thus surpassing by multiple order the typical neuronal spiking frequency (of the order of 10 Hz).
\\
(2) The amplitude threshold  is varied by less than 10 dB within the hearing frequency range of 20 dB.

As an illustration see a shifted sine wave  for low $f_s=0.001$ and high $f_s=0.5$ frequencies with $A_s=1.2$ with high white noise. The receiver with the resolution bandwidth $\mathrm{RBW}=0.1$ does not allow to capture the signals (see red lines Fig. 2d,e correspondingly). However, the same receiver can capture the signals after the FHN transformation, which integer multipliers the frequency by $32$ and $1/15$ correspondingly, adjusting it to $f_0$, which is suitable for the receiver detection.

Thus, the FHN spiking behaviour acts as a modulator, which adapts the signal frequency to the value characteristic to the FHN system and, thus, suitable for reception. This is achieved by the HFT, see Fig. 2b) (see the inset of Fig. 1d) for the bandwidth of encoded signals). The fractal nature (infinite self-similarity) allows to probe even high frequency deviations from $f_0$. Moreover, the higher the deviation  ($\Delta=f_s-f_0$) the better is the precision (output deviation $\Delta'=\Delta/[f_0/f_s]$, precision $\Delta'/f_0\simeq /[f_0/f_s]$).  Consequently, unlike other regeneration types, it enables high efficiency regeneration even for large deviations due to self-similar, scaled and infinitely repeated pattern (see Fig. 2b) centered around the single attraction point $f_0$. Thus, this is a novel regeneration type - fractal regeneration, which allows suppression of high deviations (see Supplemental Note III). Suppressing high deviations, it ensures \textit{stability} of communication networks. As an example, consider the case when signals, e.g. in Figs. 2d,e), were sent at higher or lower frequency than intended due to signal distortion or failed network node, then the FHN-augmented network will be able to sustain communication analogous to functional connectivity in brain. This is also important for 6G systems, which has high demands in frequency packing. 

Re-modulation is a fundamental nature of FHN, which acts as a nonlinear frequency-amplitude modulator (as shown by Fig. 1c), sharing similarity of typical ASK- and FSK-spectra: both the central band and the peaks at eigen-frequency are essential components of the FHN output spectrum. Integer re-modulation, arising from harmonic locking, is crucial for sampling, so that not only signal can be detected as in Figs. 2d,e), but also information retrieved. As a demonstration we consider a random encoding of $2^{25}$ symbols encoded as zeros and ones (OOK) modulated with a rectangular waveform with the stimulus amplitude $A_s=1.2$ and $R_0/f_s=2$(see input and generated spectrum in the Supplemental Note II.B.; as a demonstration of multilevel encoding see 4-level amplitude signal in Supplemental Note II.C). As a result of FHN transformation the bandwidth is increased to $B=f_s$ and the output Nyquist rate equals to $R=2f_s$, i.e. oversampling is enabled. To illustrate the impact on processing efficiency, the white noise is added to the input signal.  Here we use the same rectangular matched filter for both linear and FHN cases  with varied number of samples per symbol $N_s=2, 4, 16$ (based on the input symbol frequency) and plot the corresponding error vector magnitude (EVM) improvement over the corresponding linear oversampled case (note, both cases - nonlinear and linear - are equally oversampled, thus, highlighting the improvement due to noise shaping only) plotted in Fig. 2c). One can see that higher  improvement is achieved for the signal below the noise floor   with the maximum improvement of 10 dB at signal-to-noise ratio $SNR=-3 dB$ and $N_s=4$.  For this case in Fig. 2e) compare the output noise (i.e. when the input signal is deducted from the output) after the linear processor (shown in blue) and after the FHN nonlinear processor (shown in red):  after FHN the noise is suppressed within the signal bandwidth window (zone I in Fig. 2e)) and dispersed forming characteristic spikes at frequencies  integer multiple of $f_s$ (see zone II), thus, the pattern repeating with $f_s$ frequency (see zone III). So that, when the signal is oversampled the received samples have nonlinearly reduced reduced. 
The observed signal spectrum structure with awn-like spikes and its functionality of noise dispersing is analogous to the wheat and emmer awns and their role in seed dispersing \cite{zerno}. The topology of the HSCT is also reminiscent to the geometric shapes in musical instruments (e.g. organ and accordion). 

Overall, the bandwidth and frequency harmonic fractal transformation in Fig. 1d) and Fig. 2b), correspondingly, represents a two-fold advantage as (i) 4R- and (ii) fractal- regeneration. The first is due to re-modulation functionality, which enables a regeneration in frequency domain, similar to re-timing for suppressing temporal jitter in 3R \cite{3R1,3R2,3R3}. The second is related to a fractal functionality: previous regenerators suppressed noise within the decision boundary (e.g. Voronoi region for constellations) moving it outside, here the decision boundary is infinite but fragmented and appropriately scaled (see Supplemental Note 5): the noise suppression is achieved due to re-shaping noise in repeated patterns (see Fig. 2f). While the conventional regeneration (2R, 3R, etc.) \cite{NC} allows to suppress only small deviations (noise) \cite{May} (as large deviations can only be increased). Whereas, the fractal regeneration suppresses larger deviations stronger, while small deviations are preserved (frequency is preserved if it is of the order of eigen-frequency and transformed only if it deviates strongly). This concept can be generalized to the central-symmetric quantization of a real-valued continuous variable $x$ with regard to the quantization centre $x_0$: $C(x)=x[x_0/x]$ in contrast to the conventional quantization with the step $x_0$: $Q(x)=x_0[x/x_0]$ (see Supplemental Note III), consequently the FHN neuron can be viewed as an analogue quantizer-modulator characterized by the corresponding quantization frequency.

\section*{Discussion}
Overall, quantization is at the foundations of modern computing and communications, including sampling, compression and coding. While there is a change of paradigm in analogue  computing caused by the advancements in neural networks and artificial intelligence - neuromorphic and non-von Neumann computing - necessitated by the growing energy and complexity requirements, there is a need to revisit the foundations of communications for incorporating analogue neuromorphic approach. 

Here we illustrate using the FHN circuit - one of the simplest neuronal models - the effect of harmonic fractal frequency and bandwidth transformation, which allows to adapt and capture signals with frequencies multiple orders lower or higher the receiver resolution offering high sensitivity and stability of the network  with simultaneous noise shaping for robustness in a single simple RLC circuit. We reveal the first physical mechanism, which allows to model results analogous to the high hearing sensitivity in animals and high stability in functional connectivity in the brain. The FHN simple circuit architecture allows to achieve fragmented bandwidth dependency unlike the conventional band- or low/high-pass filters and demonstrates possibility to achieve noise shaping without  a delay line. It also unveils a novel neuromorphic approach to regeneration and quantization. 
 Our results pave the way for a paradigm change in the design of channel components, including quantization, sampling, modulation, filtering and regeneration. 

\section*{Methods}
The FHN equation is calculated numerically by the adaptive Runge-Kutta method with the variable step size (with the computational threshold of $3\%$). The spectra were obtained by using the Fast Fourier Transform with the varied spectral step to achieve accuracy above the $3\%$. The analytical approximations were obtained by adapting the Poincar\'e-Lindstedt method \cite{PL1,PL2}.

\section*{Supplemental materials}
\section{Constant stimulus}
Here we consider a constant in time stimulus $I=A_s=\mathrm{const}$ and investigate how the spike frequency and amplitude  varies as a function of $A_s$.
\subsection{Spectrum of constant stimulus}
\begin{figure*}[!ht]
\centering
\includegraphics[width=\linewidth]{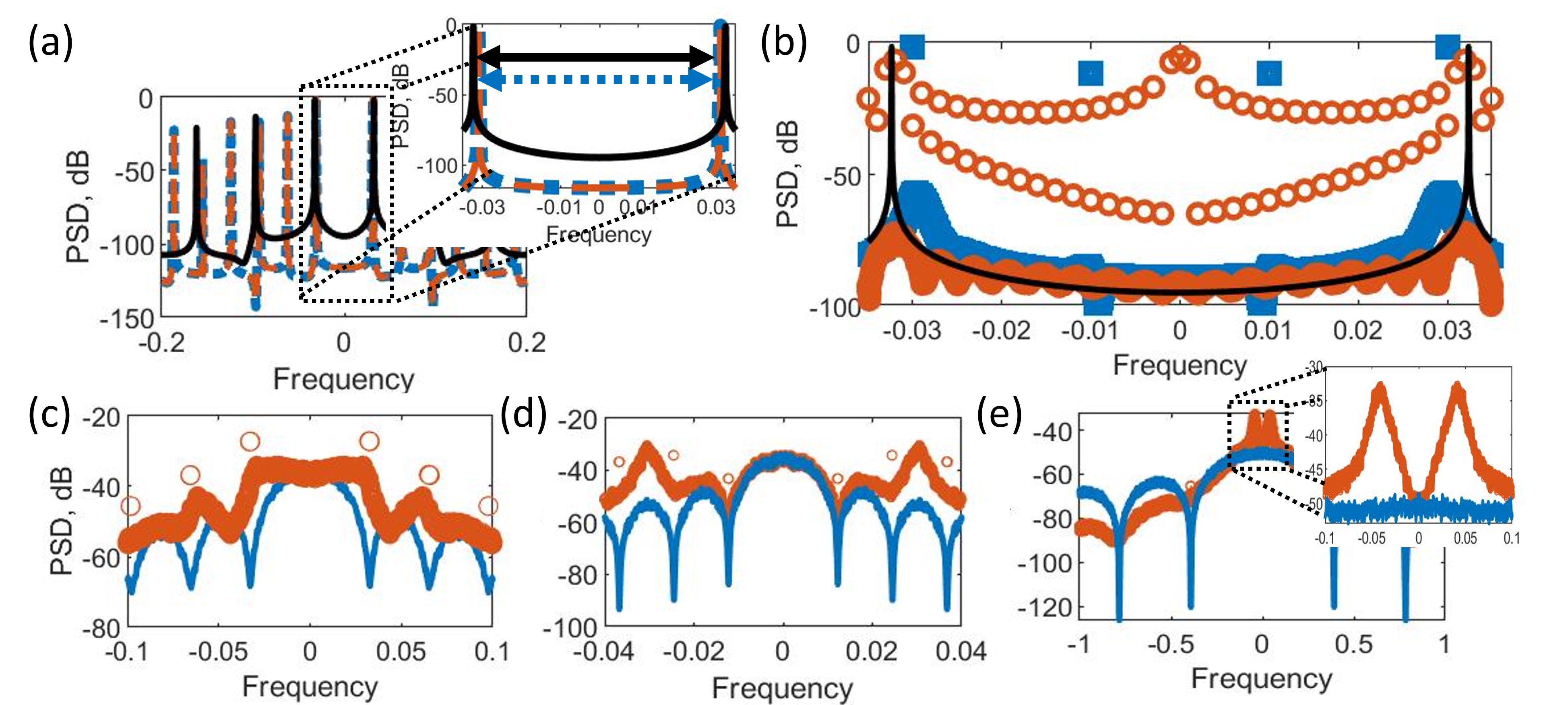}
\caption{\textbf{Spectra.}  
\textbf{a} Output spectrum and its main components (inset) for \textit{constant}  stimulus with amplitude:  $A^*=a/b$ (black, solid lines) and $A_s=0.6;1.15$ (red, dashed; yellow, dotted). The output spectra represent frequency combs: spaced at  $f_0(A_s)$ (oscillation frequency) for $A_s\neq A^*$;  the frequency and spacing is symmetric around the balancing value $A_s=A^*$, see red dashed and black solid lines for $A_s=A^*\pm 0.275=0.6;1.15$; the spacing is doubled for the balancing value $2f_0(A^*)=2f^*$ (octave). The inset demonstrates small deviations and symmetry of the frequency values $f_0(A_s)$ (see Fig. 1c) for $A_s=A^*\pm 0.275=0.6;1.15$  from the maximum value $f^*$ at $A_s=A^*$  (solid black) as highlighted by coloured dashed arrows, correspondingly.  \textbf{b} Output spectra for rectangular \textit{periodic}  stimulus  with frequency $f_s=0.001$ (red, circles) and $f_s=0.01$ (blue, squares) with the output spectrum for constant stimulus (black, solid) plotted for comparison. When the stimulus is periodic the generated signal   spectra represents an interplay of frequency combs with $f_s$ and $f_0$ spacing. Thus,  the closest harmonic ($nf_s$) to the eigen-frequency $f_0$ becomes elevated (consequently, the largest spectral components have frequency $nf_s$ with $n=[f_0/f_s]$ for $f_s<f_0$). \textbf{c-e}  In the \textit{encoded} case the output spectrum (shown in red lines) represents an interplay of a continuous input spectrum (see  blue lines) driven by $f_s$ (symbol frequency, also referred as symbol rate) and a comb driven by the eigen-frequency $f_0$ (red circles), with the corresponding eigen-rate $R_0$. As a result,  the output Nyquist rate is given by $R_{f_s}=f_s[R_0/f_s]$, see $R_0/f_s=2, \ 5.2, \  1/6$ resulting in $R_{f_s}/f_s=2, \ 5, \ 1/4$ (compare to Fig.1d)).}
\label{SMF1}
\end{figure*}

By plotting the output power spectral density (PSD) of the FHN response to a \textbf{constant} stimuli (see Fig. 3a) one can observe the formation of frequency combs which reveal the symmetry of the oscillation frequency (which also plays a role of frequency comb spacing) $f_0$ of the FHN equation to the stimulus amplitude $A_s$ deviation from the value $A^*=a/b$: $f_0(A^*+\delta A)=f_0(A^*-\delta A)$. When $A_s=A^*$ one can observe the doubling of the frequency comb spacing (i.e. in this case frequency comb spacing equals to $2f^*$).  

This shows that the FHN \textit{comb} is highly sensitive to the stimulus amplitude - and can potentially be used as a highly sensitive amplitude sensor.
At the same time, note the \textit{stability of oscillation frequency} for varied stimulus amplitude (see small deviations of oscillation frequency highlighted in the inset of Fig. 3a), which is further studied analytically).

\subsection{Analytical results for constant stimulus}
Here we consider a constant stimulus $I=A_s=\mathrm{const}$.
Firstly, we rewrite the FHN equations as:
\begin{equation}
\ddot{v}=\dot{v}\left(1-v^2-\frac{b}{\tau}\right)-\frac{v}{\tau}(1-b)-\frac{bv^3}{3\tau}+\frac{b\varepsilon}{\tau}
 \label{Eq2}
\end{equation}
with $\varepsilon=I-a/b$. This shows that at $I=A^*=a/b$, the equation is symmetric with regard to the sign inverse. Thus, for $\varepsilon=0$ the positive values of $v$ are symmetric to the negative, consequently, the  $\mathrm{max}(v)=-\mathrm{min}(v)$.  We refer to this value of the stimulus as the balanced value and denote $A^*=a/b$.
Let us assume that the solution $\{v^*, \ w^*\}$ is known for the stimulus $I=A^*$. The corresponding frequency and period will be correspondingly denoted as $f^*$ and $T^*$.

For $\varepsilon\ll 1\}$ we expand the solution in amplitude and in frequency according to the Poincar\'e-Lindstedt method. We rescale the time and the oscillation frequency as $\xi=t(1+\epsilon \zeta)$ and $f=f^*(1+\varepsilon \zeta)$. Here we focus on finding an asymptotic solution on the  frequency of oscillations rather than full solution, therefore, adapt the Poincar\'e-Lindstedt method focusing on the extremum points only. For convenience we use the following notations: $\mathrm{max}(v)=v^+; \ \mathrm{min}(v)=v^-$. Then in the main order expansion, we  receive: \[
\ddot{v}_0^{\pm}=-\frac{v_0^{\pm}}{\tau}(1-b)-\frac{b(v_0^{\pm})^3}{3\tau}
\]
$v^+=-v^-\simeq\sqrt{3}$. Similarly, for the first order expansion we receive: 
 \[
2\zeta_1\ddot{v}_0^{\pm}+\ddot{v}_1^{\pm}=-\frac{v_1^{\pm}}{\tau}(1-b)-\frac{b(v_0^{\pm})^2v_1^{\pm}}{\tau}+\frac{b}{\tau}
\]
by re-grouping terms of the same order and from the symmetry analysis, we receive $\zeta_1=0$ and $v_1^+=(v_0^+)^{-2}=1/3 $. Here we encounter asymmetry in the solution of $v$ (arising from the free term governed by $\varepsilon$) with extremum values smaller/larger (similarly, the positive branches become shorter/longer) depending on the sign of $\varepsilon$ - for the stimulus smaller/larger than the balancing value $A^*=a/b$, in particular $v$ experiences a constant shift in addition to temporal variation $v \simeq v_0 + \varepsilon + \varepsilon F(t)$ (with the zero-order solution defined by the balanced solution, i.e. $\varepsilon=0$: $v_0=v^*$) and, consequently, the spectrum receives a discrete component at zero frequency $\varepsilon\delta(f)$ (as the Fourier transformation of the constant term). Moreover, as the spike initiates and grows one needs to account for the assymetry in the extremum values obtaining:  $v_1^-=1/2-1/3=1/6$. 
In the next order:
 \[
2\zeta_2\ddot{v}_0^{\pm}+\ddot{v}_2^{\pm}=-\frac{v_2^{\pm}}{\tau}(1-b)-\frac{b(v_0^{\pm})^2v_2^{\pm}}{\tau}-\frac{b(v_1^{\pm})^2v_0^{\pm}}{\tau}
\]
from the symmetry analysis and comparing terms of the same order we receive $\zeta_2=-1/2$ and $v_2=-(v_0^{\pm})^{-5}$. 

Thus, we receive the asymptotic expansion for the frequency of spikes when driven by unmodulated stimulus: $f=f^*[1-(I-a/b)^2/2]$ with $f^*$ - a frequency of the spikes when driven by the stimulus $A^*=a/b$.  

Thus, we obtain the asymptotic expression for the output signalling frequency: $f_0=f^*(1-(I-a/b)^2/2)$ with $f^*$ - a frequency of the spikes when driven by the stimulus $I=A^*=a/b$ ($f^*$ can be well approximated as $(\pi\tau)^{-1}$ for large $\tau$). See  numerical results (in bold) for various $tau$ values (and fixed $a=0.8, b=0.7$) as a function of stimulus amplitude compared with the derived analytical approximation (in circles) in Fig. 4a).
The derived asymptotic expressions (shown in circles)  for the extrema  ($\max(v(A))=\max(v(A^*))+(A-A^*)/3$ and  $\min(v(A))=\min(v(A^*))+(A-A^*/6$)  are plotted in Fig. 4b,c)  alongside the numerically obtained values (solid lines).

\begin{figure*}[!ht]
\centering
\includegraphics[width=\linewidth]{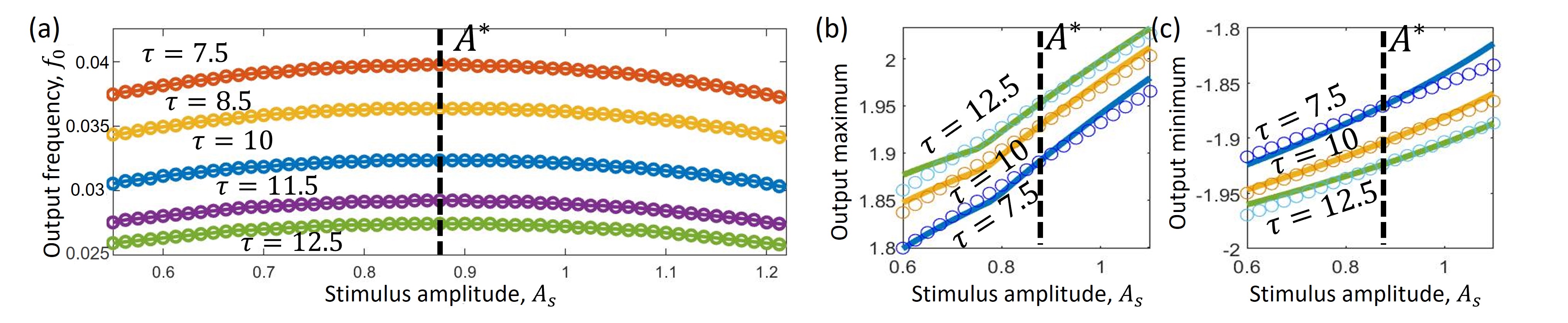}
\caption{\textbf{Constant stimulus.} The output spike \textbf{a} frequency and \textbf{b,c} amplitude extrema as a function of stimulus amplitude  for varied FHN parameter $\tau$. The balancing amplitude $A^*$ is highlighted by the vertical line.  Here numerical and analytical approximations are shown by solid lines and circles, correspondingly. The spike frequency for a constant stimulus represent a characteristic spiking range at which the FHN neuron operates - eigen-frequencies.   } 
\label{SMF2}
\end{figure*}

\section{Modulated stimulus}
Here we incorporate stimulus frequency: we consider a stimulus modulated with frequency $f_s$: $I=A_sP(f_st)$ and investigate how the spike frequency and amplitude  varies as a function of $A_s$ and $f_s$. In A) we consider a periodic (uncoded) case - a square wave $I=A_s\mathrm{sgn}\sin(2\pi f_st)$ ($\mathrm{sgn}$ denotes a sign function) and derive the corresponding analytical results; in B) we consider an encoded case -  $I=A_sc_k \Pi(f_st)$ ($\Pi$ denotes a square pulse) and derive the corresponding analytical results, we also we consider an encoded case for other waveforms: sinc and rrc and varied parameters.

\begin{figure*}[!ht]
\centering
\includegraphics[width=\linewidth]{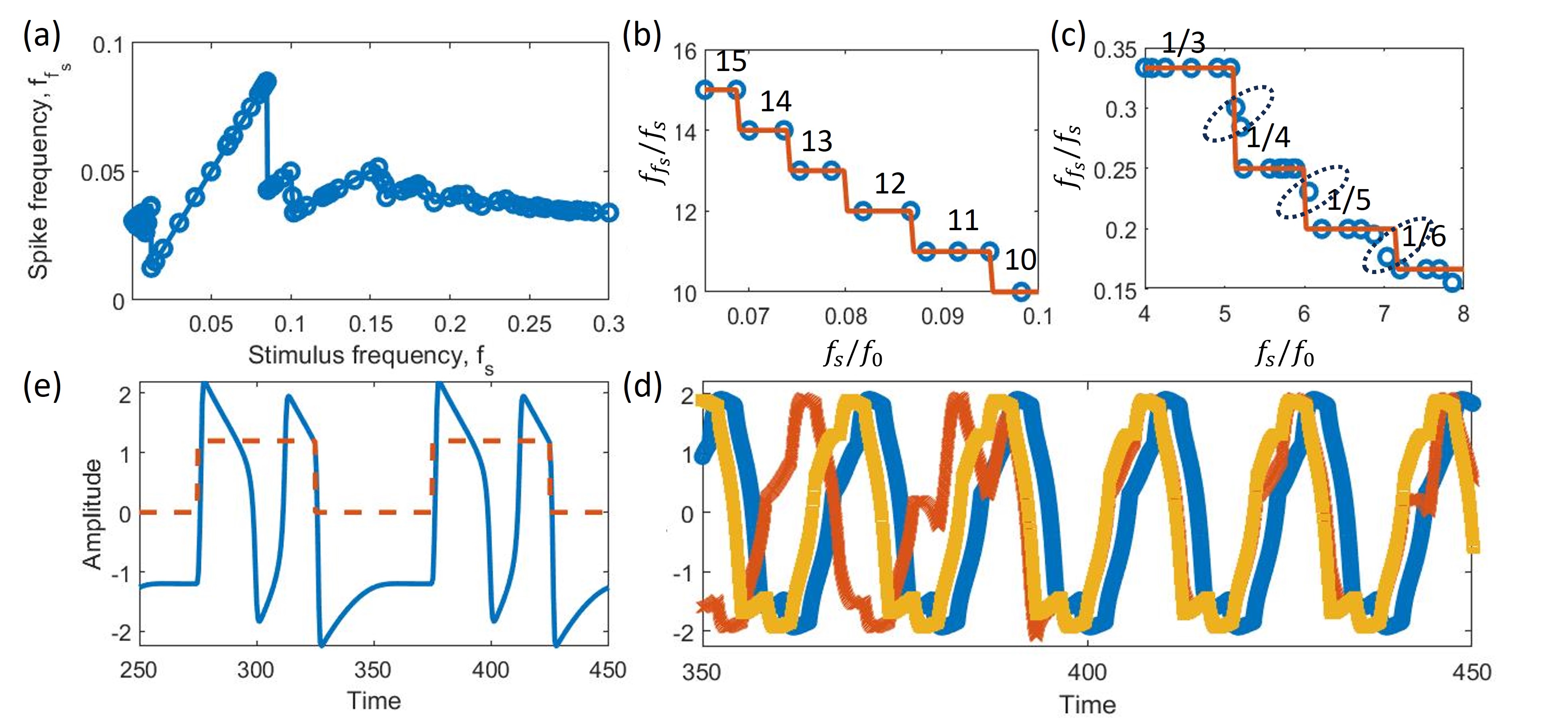}
\caption{\textbf{The spike frequency:}  \textbf{a} $f_{out}=f_{f_s}$ as a function of  stimulus frequency $f_s$ (the amplitude is fixed to $A_s=1.2$).  The spike frequency is preserved $f_{f_s}\rightarrow f_0$ to the corresponding eigen-frequency $f_0$ (the spike frequency excited by a constant stimuli with the corresponding amplitude $A_s$, see Fig. 4a); \textbf{b,c}  the spike frequency ratio $f_{f_s}/f_s$ to the input frequency for within-the-threshold stimulus (here $A_s=1.2$) is an \textit{integer} \textbf{b} multiple (for $f_s<f_0$) and \textbf{c} submultiple for $f_s>f_0$ of the stimulus frequency $f_s$.  For $f_s>f_0$ the spike frequency is the inverse of the output period: $f_{f_s>f_0}=1/T_{out}$, for $f_s<f_0$ the bursting regime takes place: multiple spikes per burst with the spike frequency $f_{f_s>f_0}>1/T_{out}$.  For high  frequency stimulus the waveform plays a crucial role and may lead to the violations from the phase locking (as highlighted in circles). \textbf{d} Compare the output for the rectangular input with $A_s=1.2$ and frequencies $f_s=0.156$ (blue) and $0.157$ (red). The first case exhibit a regular spiking behaviour and phase locking $T_{out}=3T_{in}$ with $f_{f_s=0.156}=1/3f_s$, whereas the second example results in irregular spiking with the spike frequency  $f_{f_s=0.157}=0.3f_s$. Note, the same frequency with different amplitudes will exhibit a behaviour ($f_{f_s=0.157,A_s=1.25}=1/3f_s$ and $f_{f_s=0.157,A_s=1.15}=1/4f_s$, correspondingly), the latter case is plotted in yellow. \textbf{e} For $f_s<f_0$ the spikes are generated with frequency  $f_{f_s}$ integer multiple of $f_s$, see the input with $f_s=0.01$ and $A_s=1.2$  in red, while the corresponding output spiking frequency is integer multiplied: $f_{f_s=0.01}=3f_s$. }
\label{SMF3}
\end{figure*}
\subsection{Periodic stimulus}

When driven by a periodic slow stimulus (see Fig. 3b) the interplay of the combs with spacing of the stimulus frequency  $f_s$ and eigen-frequency $f_0$, resulting in the component $nf_s$ closest to $f_0$: $n=[f_0/f_s]$ to be elevated. This has a meaning of the spike frequency: when looking at the Fig.1b) one can see that the input pulse becomes split into a burst consisting of multiple pulses, while the frequency of the burst is preserved to the stimulus frequency $f_s$, the output spike frequency $f_{f_s}=f_s[f_0/f_s]$. See the numerical values of the output spike frequency $f_{f_s}$ in see Fig.5a) and the numerical value $f_{f_s}/f_s$ (in blue circles)  with the analytical formula $n$ with $n=[f_0/f_s]$ (red solid lines) in Fig. 5b). 

For comparable  frequencies $f_s\simeq f_0$, the elevated component is $n=[f_0/f_s]=1$ and the output pulse frequency equals to the stimulus (there is one generated pulse for each input pulse).

For a very fast stimulus $f_s/f_0 \rightarrow \infty$ the FHN response time (mainly governed by $\tau$), i.e. the time required to transit trajectories in the phase space, is to small for the excitation dynamic, and the system saturates to treating the input as a constant excitation with averaged amplitude $A_s \rightarrow \langle A \rangle$, thus, the output frequency saturates asymptotically to the output frequency $f_0$ in constant stimuli case: 
\[\lim_{f_s/f_0 \rightarrow \infty} f_{f_s} =f_0(\langle A \rangle)= \hat{f}\] (see Fig. 5a,c)). 
For a fast stimulus, $f_s/f_0 \ll 1$ a similar dynamics to Fig. 3b) is observed, again there is an interplay of two combs with spacing $f_s$ and $f_0$, however, as in this case $f_s > f_0$, then the component 
\[f_{f_s> f_0, \mathrm{not} f_s\gg f_0}=\frac{f_s}{[f_s/f_0]}\] is in resonance and becomes elevated. These two  asymptotic expressions for a fast and a saturating case  can be combined as:
\[f_{f_s > f_0}=\frac{f_s}{\Big[\frac{f_s}{f_0(1+1/[f_s/\hat{f}]}\Big]}\]
which incorporates the above two cases and is plotted in red lines in Fig. 5c) alongside the numerical results plotted in blue circles.

For very high stimulus frequencies $f_s\ll f_0$ the deviations from the above analytical formula are observed at the narrow frequency intervals in-between the stair-case harmonic locking. As described above for $f_s/f_0 /\rightarrow \infty$  there are two governing frequencies in the system defined by the waveform and amplitude:  $f_0(\langle A \rangle)$ and $f_0(A_s=\max(A))$, which creates additional regime types. For example, it may lead to exceptional sensitivity to the stimulus waveform, amplitude  and/or frequency: in Fig. 5d) we plot three outputs for nearly identical parameters of the rectangular stimulus, which demonstrate how the small difference in amplitude or frequency may violate the regularity of excitation. This is important as depending on the waveform at high frequencies one may observe an excitation block even at the values well within-the-threshold for the constant stimulus. The role of stimulus waveform is less pronounced for small frequencies, however it may be crucial for randomly encoded signals. 

For small frequencies it is important to note the difference between the burst period and spike (i.e. pulse) period. In Fig. 5e) we plot an input stimulus (in red dashed lines) with frequency $f_s=0.01$. This generates bursts (shown in blue solid) with output period equal to the input period: $T_{out}=T_{in}=1/f_s$, while the spike frequency can be measured  via the spectrogram and results in $f_{f_s=0.01}=3f_s$ in agreement with the analytical derivations.

\subsection{Encoded stimulus.}
Similar results we observe for modulated signals. We remind that the encoded signal is \textbf{non-periodic}, therefore, the \textbf{spectrum is continuous}. See, input signal spectra for a rectangular waveform with amplitude $A_s=1.2$ at Figs. 3(c-e)  shown in blue lines for various symbol frequencies. The resulted FHN spectrum is plotted in red circles, so that arising from the FHN transformation discrete components are clearly shown - one can see that the discrete components (red circles) arise at the corresponding frequencies - integer multiple of $f_s$, moreover, the spikes appear at frequencies in the vicinity  to $f_0$ - preserving bandwidth to the same value - \textbf{eigen-bandwidth} independently of input stimulus frequency (or waveform as demonstrated in Fig. 6).

We further consider  sinc and raised cosine (rrc), which have different spectrum and bandwidth. The ideal realistic sinc pulse has a finite spectrum, i.e. PSD changes from $-f_s/2$ to $f_s/2$, and, therefore, sinc pulses have a minimum bandwidth among other waveforms for a given symbol frequency $f_s$: $f_s/2$ and, consequently, the minimum Nyquist rate: $f_s$. Sinc pulses can be realised in practice with the corresponding sinc filter cut in temporal domain at the finite length measured in symbol period - number of taps: compare spectra at 10 and 100 taps in Fig. 6a,b). This leads to the slight increase in bandwidth. In contrast, see rrc with 0.5 rolloff in Fig. 6c) with excess bandwidth $50\%$ compared to the ideal sinc pulse with the same $f_s$. Nevertheless, one can see that in all examples the discrete components (red circles) arise at frequencies multiple of $f_s$ (highlighted by grey solid lines in Fig. 6) independently of the waveform or bandwidth, depending on symbol frequency only - $f_s$. The spikes also arise at the same value $[R_0/f_s]$ independently of the waveform or bandwidth (as highlighted by black dashed lines in Fig. 6).

It is interesting, that the spike width, although being very sharp (i.e. compared to the signal input spectrum width), depends on the waveform (compare Figs. 3d) and Fig. 6 having the same $f_s$ and $A_s$) and become more narrow for larger bandwidth. 
Note, that although  in Fig. 3d) the input spectrum is naturally modulated with frequency $f_s$, this is not the case for the spectra in Fig. 6. This illustrates that the discrete components in the output of FHN arise from periodicity of the coding scheme.

Now, let us study in more detail the impact of encoding, i.e. having input continuous spectra. We return to the rectangular waveform spectra in Fig. 3 (c-e) and consider the impact of $f_s$ variation.  For medium values of symbol frequency  $f_s\simeq R_0$ (see Fig. 3c) for $R_0/f_s$=2.4) the main lobes  merge with the spiked lobes preserving the bandwidth to the eigen-value and integer multiplying the Nyquist rate (here $R_{f_s}=2f_s$). Whereas for small values of $f_s\ll R_0$ (here $R_0/f_s=5.2$)  the main lobe is preserved as in the stimulus, while the spikes emerge at the $\pm f_0$ (see discrete components in red circles). As encoded signal spectrum has a continuous part (unlike uncoded periodic case, compare to Fig. 3b)) an additional spike in-between - in the middle between the discrete components - is possible (two neighbouring spectral components may act in resonance, if they are symmetric around $f_0$, elevating the central part between them), thus the peak is formed at the value $n=[2f_0/f_s]=[R_0/f_s]$)  (here $R=5f_s$). For large values of $f_s\ll R_0$ (here $f_s=1/6 R_0$ with the resulting output bandwidth $R=1/4 f_s$) the main lobe becomes split creating spikes close to $\pm f_0$ (see inset). In all cases the signal is transformed so that the output Nyquist  rate is the \textbf{integer} multiple or submultiple of the input stimulus frequency.

\begin{figure}[!ht]
\centering
\includegraphics[width=0.7\linewidth]{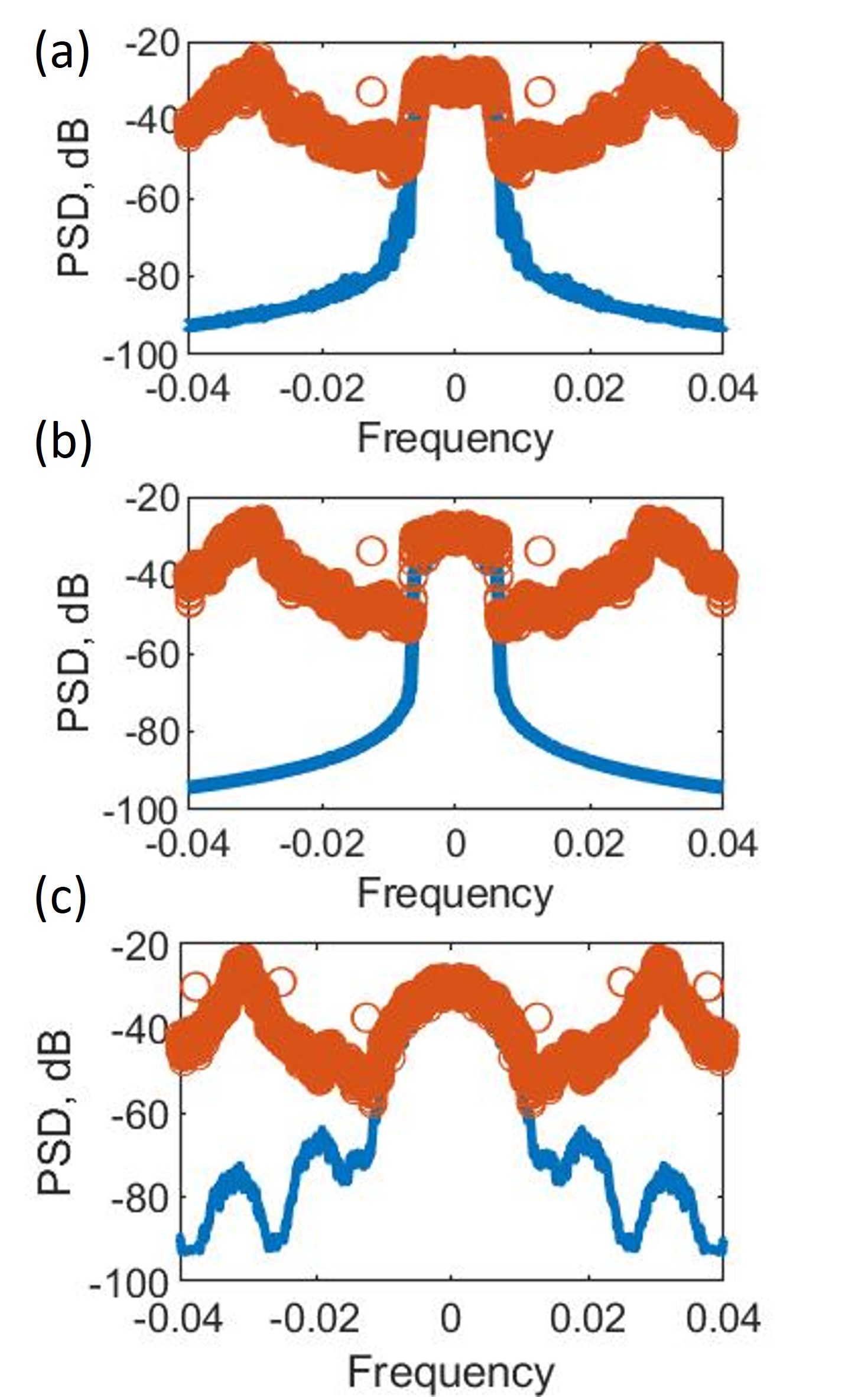}
\caption{\textbf{Spectra of encoded signals with various waveforms.}  
 The output spectrum (shown in red lines) and its input spectrum (see  blue lines) for FHN driven by the same symbol frequency $R_0/f_s=5.2$ for various modulation waveforms: \textbf{a} sinc-pulse with 10 taps (i.e. small filter memory and small irregularity in spectrum), (b) sinc pulse with 100 taps (approximation of an ideal sinc pulse), (c) raised cosine with $100 \%$ roll-off with 1 tap (i.e. $100 \%$ higher bandwidth compared to the input in panel b) and no filter memory). The output spectrum is an interplay of discrete components (shown by red circles) at frequencies integer multiple of $f_s$ and a continuous part, which repeats the input at small frequencies and exhibit spikes at frequency integer multiple of $f_s/2$ closest to $f_0$ resulting in integer multiplication of the output Nyquist rate (here $R=5f_s$) for all waveform cases. 
}
\label{SMF4}
\end{figure}
\subsection{4-level encoded stimulus}
The effect of noise shaping shown in Fig. 2c,f) can be observed in various complex encoding schemes as a simple demonstration we consider 4-level amplitude encoding: we keep the same waveform parameters changing only the coding scheme. So, the symbols $c_k$ are now randomly drawn from the alphabet $\{0, 1, 2, 3\}$. The signal is then normalized keeping the same maximum amplitude $A_s=1.2$. Similarly, we add white noise at $SNR=-3 dB$ and process a linear and an FHN processed signals  with a rectangular matched filter with oversampling $SpS=4$. The output of the FHN is plotted in red in Fig. 7a) alongside the input clean signal (black), while the constellation diagrams are plotted in Fig. 7b): compare output constellations without (linear case in blue) and with (red) the FHN processing (the input constellations are shown for the reference in black). For convenience, we show separately constellations for each coding level in the insets.  
\begin{figure}[!ht]
\centering
\includegraphics[width=\linewidth]{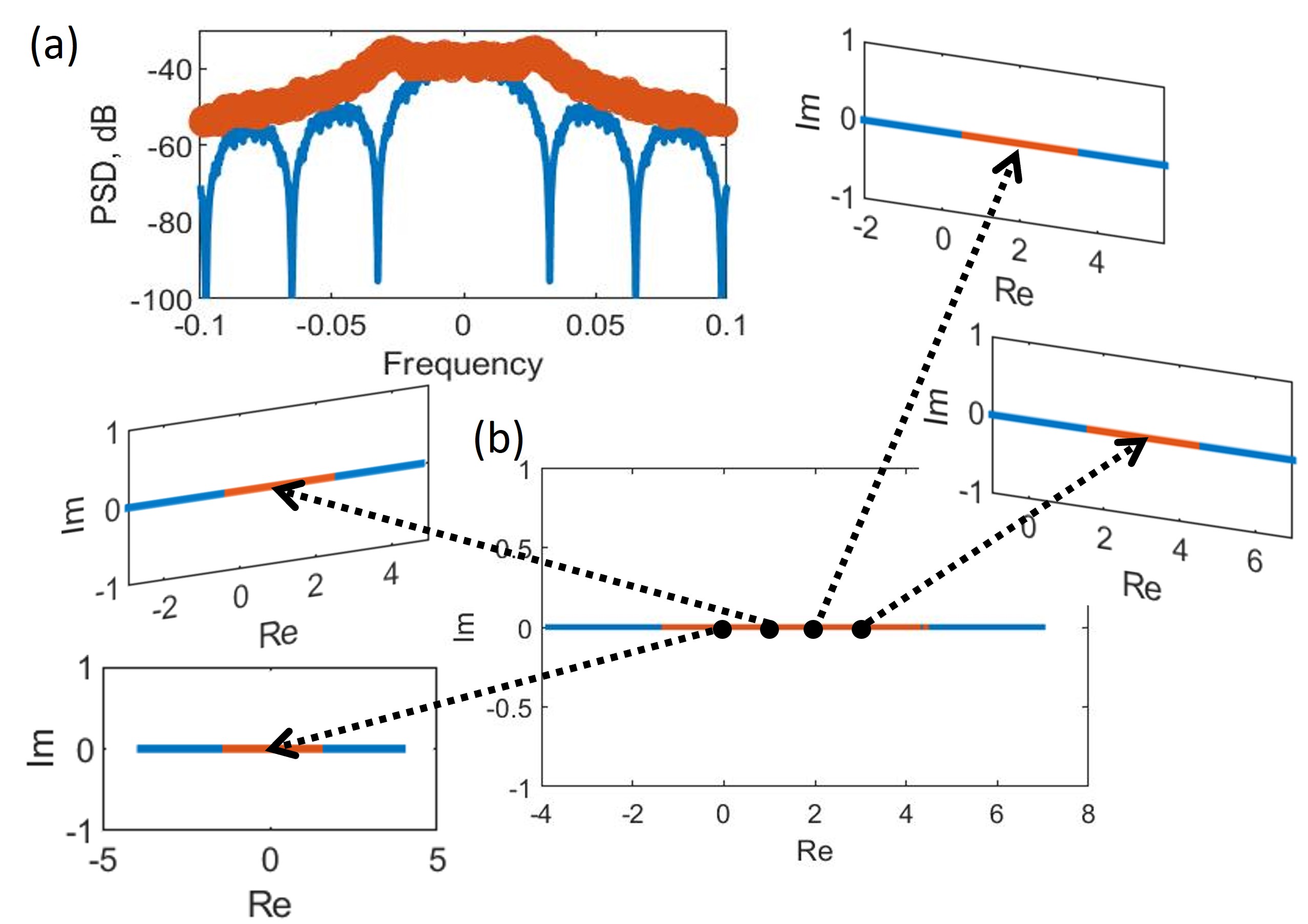}
\caption{\textbf{Spectrum and constellation diagram of a signal with four amplitude level encoding.}  
\textbf{a} The output spectrum (shown in red) and  input spectrum (shown in black) for FHN driven by the same symbol frequency $R_0/f_s=5.2$ with rectangular waveform for \textbf{b} four level constellation shown by black circles, compare a linear processing  (blue) with FHN processing (red), both with the same oversampling factor $SpS=4$; the constellation diagrams for each constellation symbol are shown in the insets correspondingly. The noise shaping (see panel a) enables an efficient noise suppression for each constellation symbol as demonstrated in panel b and the insets.
}
\label{SMF5}
\end{figure}
\section{Regeneration: design rules.} 
To achieve 4R regeneration with the threshold we augment the conditions for deviation suppression \cite{NC}, here the condition for bandwidth regeneration  starting with unmodulated case $f_s=0$ looking for the point(s) $A^*$, which satisfy
\begin{equation}
B'_{f_s=0}(A^*)\leq 1
\end{equation}
The corresponding point $A^*$ and the condition (i) is shown in Fig. \ref{F1}d). Also, we have
(ii) he condition of amplitude threshold (highlighted in Fig. \ref{F1}d), which effectively filters sampling points with low or high amplitude (see for illustration Fig. \ref{F1}b). 
Under these conditions an output of modulated stimulus converges to the stable bandwidth value $B^*$ as deviations for various amplitude samples are effectively suppressed (compare a step-like responses in the modulated and unmodulated cases in Fig. \ref{F1}d)). As a result, we receive plateau response for output bandwidth in a modulated regime, see Fig. \ref{F1}e). For comparable frequency values $f_s\simeq f_0$, the FHN processor preserves the bandwidth of the input stimulus (linear dependency). Whereas when further increasing stimulus frequency  for the case when input domain is close to the amplitude threshold, we again receive convergence to the constant stimulus scenario.
Thus, we receive bandwidth regeneration which activates only for frequencies or bandwidth significantly different to the characteristic frequency $f_0$, i.e.  preserving bandwidth for frequencies close to $f_0$ and  regenerating otherwise.

\begin{figure}[!ht]
\centering
\includegraphics[width=\linewidth]{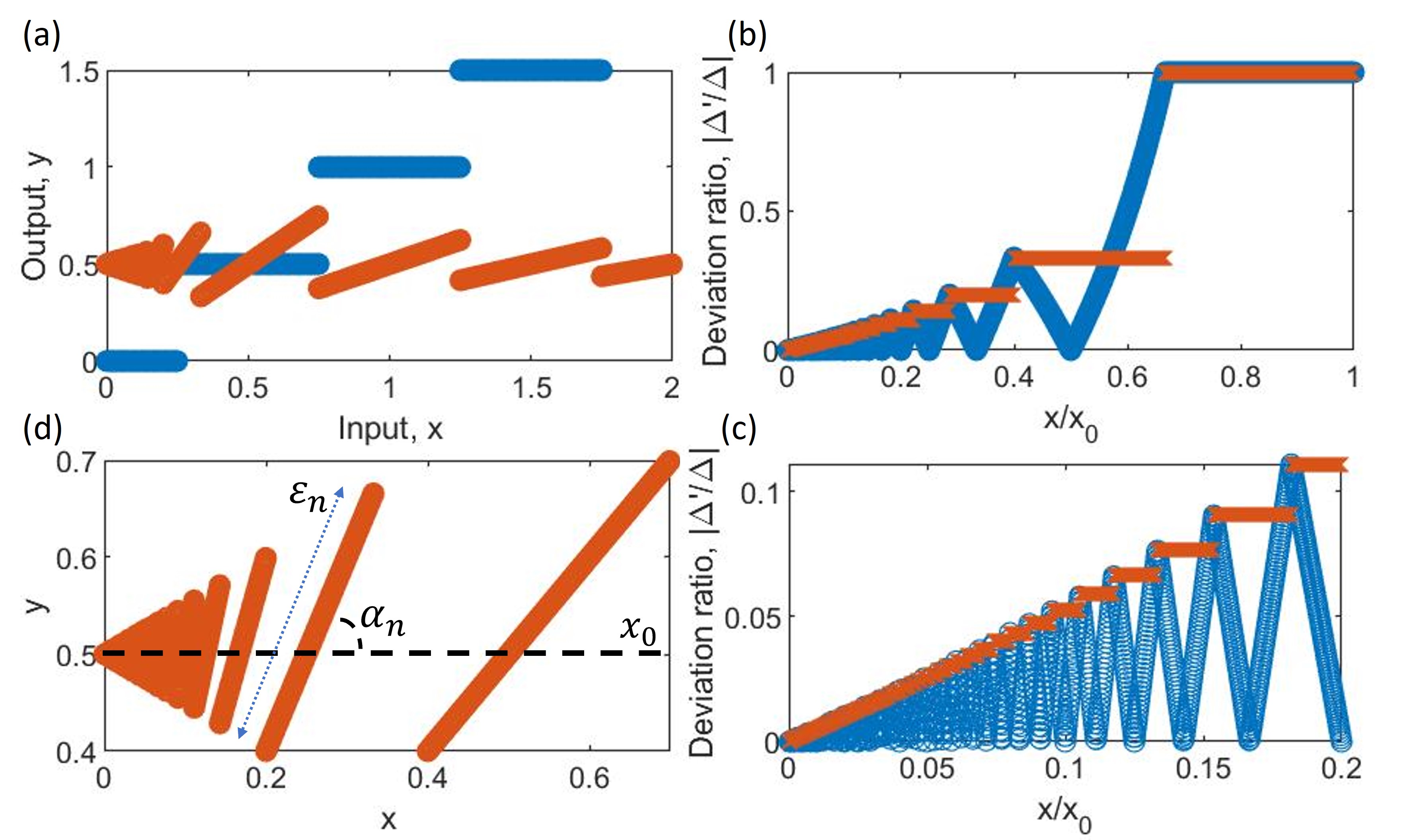}
\caption{\textbf{Harmonic fractal transformation.}  
\textbf{a} The transfer function: $H(x)=x[x0/x]$ (in red) represents a semi-analogue quantization: reduced uncertainity (deviation) around the single level $x_0$ , compare with the plotted alongside standard quantization function $Q(x)=x_0[x/x0]$ (in blue): full discretization around multiple discrete levels. 
\textbf{b} The deviation suppression $\Delta'/\Delta$ (blue) and its lower bound (red), enlarged in \textbf{c}. The higher is the deviation of $x$ from $x_0$ the better is the precision of preservation to $x_0$.
\textbf{d} The fractal nature of HFT: self-similar and infinitely repeated pattern with scaling $\varepsilon_n\rightarrow x_0n^{-1}$ and angle $\alpha_n=\mathrm{atan}(n)$.
centered at $x_0$}
\label{SMF6}
\end{figure}

\section{Harmonic fractal transformation} 
First of all, we define the harmonic fractal transformation (HFT)  as: \[H(x)=x\Big[\frac{x_0}{x}\Big]\eta(x_0-x)+\frac{x}{\Big[\frac{x}{x_0}\Big]}\eta(x-x_0)\] (with $\eta$ being the Heaviside function), following the asymptotics derived in the main text and Supplemental Note I, and plot it in Fig. 8a) in red lines for $x_0=0.5$.
The structure of HFT (for example, for $x<x_0$: $H(x)=x[\frac{x_0}{x}]$) is reminiscent of the standard quantization function: $Q(x)=x_0[\frac{x}{x_0}]$ (also plotted in Fig. 8b) in blue lines). One can see that $H(x)$ has semi-analogue nature - it quantizes the input around the center of the transformation, yet leaves uncertainity (deviations) proportional to the input value $x$. In contrast, the standard quantization discretizes the input completely, yet around multiple levels. In this case there is a significant loss of information about the input. 

Further, let us consider an input $x$ and denote its deviation from $x_0$ as $\Delta=x-x_0$ (for simplicity assume $x<x_0$). The maximum output deviation occurs between the decision regions, i.e. when $x_0=(n\pm1/2)x$ ($n=[x_0/x]$), then the minimum deviation suppression 
\[\frac{\Delta'}{\Delta}=\frac{1}{2(n-3/2)};\frac{1}{2(n-1/2)}\]
for $\pm$ cases correspondingly. For the small $n$ the above result can be approximated well by $1/n$ (see Fig. 2b). Overall, the lower bound of the deviation suppression is given by $1/(2(n-1/2))$, which is plotted in red in Fig. 8b) alongside numerically calculated deviation suppression values $\Delta'/\Delta$ plotted in blue (here we leave the linear scale in the plot to demonstrate the dependence of the suppression on the harmonic $n=[x_0/x]$, whereas the enlarged interval is shown in Fig. 8c)). One can see, that unlike other regeneration types here the suppression is stronger the bigger is the deviation. 

The effect of higher suppression for bigger deviation is an important advantage of the transformation, it arises from the fractal nature of $H(x)$. The transformation not only quantizes the linear function splitting it in segments as the standard quantization function (see steps in Fig. 8a) in blue), but also turns and scales (see Fig. 8d). For $x<x_0$ for each $n$-th segment $n=[x/x_0]$ the segment length is $\varepsilon_n=\sqrt{n^2+1}x_0(n^2-1/4)^{-1}\rightarrow x_0n^{-1}$ (which asymptotically tends to $x_0n^{-1}$ for large $n$) and the angle is given by: $\alpha_n=\mathrm{atan}(n)$. The turn and scale quantization around the fixed center allows to receive a fractal type quantization: infinite self-similar transformation. 

Thus, the HFT enables a new type of semi-analogue quantization: discretization at the single level with an uncertainty at the infinitely diminishing self-similar scale.

\end{document}

%% file: preamble.tex
\usepackage{amsthm}
\usepackage{mathtools}
\usepackage{physics}
\usepackage{xcolor}
\usepackage{graphicx}
\usepackage[left=23mm,right=13mm,top=35mm,columnsep=15pt]{geometry} 
\usepackage{adjustbox}
\usepackage{placeins}
\usepackage[T1]{fontenc}
\usepackage{lipsum}
\usepackage{csquotes}